\documentclass[
preprint,
prd,
aps,
showkeys,
eqsecnum,
amsmath,
amssymb,
]{revtex4-1}

\def\Tr{\mbox{Tr}\,}

\usepackage{ifpdf}
\usepackage{cancel}
\usepackage{graphicx}
\usepackage{xcolor}
\usepackage{float}
\usepackage{subfloat}
\usepackage{subfigure}
\usepackage{slashed}

\usepackage{ccaption}
\usepackage{caption}

\usepackage{hyperref}
\usepackage[titletoc]{appendix}

\begin{document}

\title{Is Higgsium a possibility in 2HDMs? } 

\author{Ambalika Biswas}\email{ani73biswas@gmail.com}\affiliation{Vivekananda College, 269, Diamond Harbour Road, Thakurpukur, Kolkata 700063, INDIA}


\begin{abstract} 
We investigate the possibility of a Higgs-Higgs bound state in the two Higgs doublet model. Specifically we look for the effect of dimension six operators, generated by new physics at a scale of a few TeV, on the self-couplings of the heavy CP even scalar field in the model.  Following the pioneering work of Grinstein and Trott~\cite{Grinstein:2007iv}, we construct an effective field theory formalism to examine the physics of the Higgs sector. The magnitudes of the attractive and repulsive coupling strengths are compared to estimate the possibility of the formation of the $H-H$ bound state. Another way to check if a bound state is formed or not is from the formation and decay times of the bound state. The possibilities in various types of two Higgs doublet models have been discussed elaborately in the paper.

\end{abstract}

\maketitle

\section{Introduction}
Though an extraordinarily successful model, the Standard Model of elementary particle physics has been unable to 
find satisfactory answers to a few questions. Two of them involve the \textit{hierarchy}~\cite{Gunion:1989we, Hooft:1979,Haber:1985,Einhorn:1991, Jouadi:2008, Jegerlehner:2013, Jegerlehner:2018} and the \textit{triviality}~\cite{Gunion:1989we, Einhorn:1991, Sher:1989, Wilson:1971, Christiansen:2017} problems. Any theory which addresses physics beyond the presently accessible energy scales may be expected to 
predict the masses and couplings which are not predicted by the Standard Model. This poses a problem for 
the mass of the Higgs boson, as it is not protected by any symmetry unlike e.g. the intermediate 
weak gauge bosons. 
The radiative corrections to the mass of the Higgs boson are quadratic in the cutoff scale, so new physics at high 
energy scale would have a divergent effect on the Higgs mass. 
Put another way, there is no clear explanation of why the mass of the Higgs particle is not as large
as the scale of any new physics, which may be $\sim 10^{16}$GeV (GUT scale) or even 
$\sim 10^{19}$GeV (Planck scale). This is known as the \textit{hierarchy} problem. 
Since the divergences 
are absorbed into the redefinition of masses and couplings at low energy without any observable effect this can be considered as more of an academic problem.
However, at the weak scale radiative corrections at the cutoff scale and tree level mass of the Higgs field must cancel 
to a high degree of precision. One  way 
out of this situation might be new physics at intermediate energies, whose symmetries could protect the Higgs
mass from corrections at higher energy scales. \\

Coming to the issue of quantum triviality there are strong evidences in support of the idea that a field theory involving only a scalar Higgs boson is trivial in four space-time 
dimensions, but the situation for realistic models including other particles in addition to the Higgs 
boson is not  known in general. Nevertheless since the Higgs boson plays a central role in the Standard Model, 
the question of triviality in Higgs models is of great importance. 
Thus we need to look beyond the Standard Model and introduce new fields at a higher scale 
($\cal{M}\sim $ 1TeV). The simplest extension of the Standard Model is the two Higgs doublet model (2HDM)
where an additional Higgs doublet is introduced in addition to the Standard Model Higgs 
doublet (for a review see~\cite{Branco:2012}). Moreover the extended scalar sector provides 
scope to address the strong CP problem~\cite{Peccei:1977,Peccei:1977a}, matter-antimatter 
asymmetry of the universe~\cite{Turok:1991} 
and provides viable dark matter candidates~\cite{Ma:2006,Ma:2008,Barbieri:2006}.

In this paper we consider 2HDMs with a softly broken global U(1) 
symmetry~\cite{Peccei:1977,Peccei:1977a,Turok:1991,Ma:2006,Ma:2008,Barbieri:2006, Ferreira:2009}, 
with the parameters so chosen as to make the 2HDM “SM-like.”
An approximate custodial $SU(2)_{C}$ symmetry~\cite{Susskind:1979,Weinberg:1979,Sikivie:1980} has also been imposed on the SM Lagrangian density and its higher dimensional extension. This $SU(2)_{C}$ custodial symmetry must be respected by the total Lagrangian density. Also this custodial symmetry must be respected up to hypercharge and Yukawa coupling violations. Obviously there will be operators that break the custodial symmetry but in order to preserve this approximate custodial symmetry their coefficients are taken to be naturally suppressed.

The introduction of 
a second Higgs doublet and its higher dimensional extension modify the relation $M_{W} = M_{Z}\cos\theta_{W}\,,$ which is commonly parametrized by the $\rho$ parameter. The $\rho$ parameter is defined as $\rho = \frac{M_W^2}{M_Z^2 \cos^2\theta_W}$\,. This relationship is expected to be respected as precisely as possible. Since the PDG quotes $\rho_{0} = 1.00039 \pm 0.00019$ 
for the global fit~\cite{PDG:2018} of precision electro-weak observables any physics beyond the Standard Model must keep the $\rho$ 
parameter within these limits, when the particles of the new physics are integrated out. 
%
%
As we know in order to integrate out any particle from the theory it's mass must be sufficiently higher than the scale of the electroweak symmetry breaking ($v\sim 246$ GeV) thus we have to choose the new physics scale well above the electroweak scale. Thus we choose the new scale above a scale of $\cal M$ $\sim$ 1TeV. At this high energy scale the quanta of the unknown new Physics may be integrated out and we are left with a low energy effective theory. This low energy manifestation is the 2HDM supplemented with non-renormalizable local operators, of dimension $D>4$, which are 
constructed of 2HDM fields and obey the symmetry of the 2HDM. This approach is model independent, but the new physics is parametrized in terms of several arbitrary parameters and nothing is known a priori about these coefficients.

There are experimental constraints on the scale of $\cal M$. The $K^{0}-\bar{K}^{0}$ mixing restrict $\cal M$ to be $\geq 10^{4}$ TeV. Here the fact that flavour changing 
neutral currents are absent in nature has been taken into account. The Minimal Flavour Violation hypothesis~\cite{Dugan:1985,Chivukula:1987,Hall:1990,Ambrosio:2002,Cirigliano:2005,
Ali:1999,Buras:2001,Bobeth:2002,Buras:2003,Branco:2006} relaxes the bounds on ${\cal M}$ and restricts the higher dimensional operator basis. Thus a safe choice for the new physics scale would be a few TeV while naturally avoiding flavour changing 
neutral currents. 

In this paper we address the question of whether a bound state of the CP-even heavy Higgs particle can form or not. The repulsive interaction of the quartic coupling and the attractive interaction determined by the cubic coupling compete to form the bound state. For large enough coupling the exchange interaction is strong enough to produce binding. We aim to find a necessary condition on the coupling for which a non relativistic (NR) bound state may form. For this purpose we follow the procedure adopted by Grinstein and Trott in~\cite{Grinstein:2007iv} where they have formulated a non-relativistic effective theory for Higgs-Higgs interactions to study the Higgs-Higgs bound state of the SM Higgs. We borrow the name \textit{`Higgsium'} for the Higgs-Higgs bound state from their work. There the bound state of the SM Higgs particle was given this name, and it was found that this state was not likely to form for the light Higgs
particle. In 2HDM the lighter CP-even scalar $h$ is identified with the SM Higgs in the alignment limit. Here we 
attempt to find the bound state of $H$\,, which is the heavier CP-even scalar. Since $h$\, is identified with the SM 
Higgs particle, it is not likely to form a bound state, so we have adopted the same nomenclature for the $H-H$ bound state. Another way to check if a bound state is formed or not is from the comparison of the formation and decay times of the bound state. The rest of the paper has been devoted to study the possibility of the formation of the bound state in various types of two Higgs doublet models.

Formation of bound states of Higgs bosons has been discussed in the literature 
for quite some time. Much before the discovery of the Higgs boson at the LHC, 
the formation of two Higgs bound state in the Higgs model, or equivalently in 
the Higgs sector of the minimal Standard Model, has been investigated using 
different methods. The results obtained were interesting, but failed to be 
consistent with the Standard Model Higgs boson when it was actually discovered. 


In the $N/D$ method~\cite{Chew:1960,Cahn:1984} used to calculate the bound state of the Higgs boson in the Standard Model, 
the elastic scattering amplitude is written as $N(s)/D(s)$ where $N(s)$ has only left hand cuts and $D(s)$ has only right 
hand singularities. $N(s)$ is approximated by the Born amplitude which is the appropriate $s$-wave projection of the sum of the four point scattering amplitudes of the particle in picture which forms the bound state. Bound states for 
$s$-wave occur when $D(s)=s$\,, for $0<s<4m_{H}^{2}$\,. The $N/D$ method had been studied to account for the bound state
of two particles (not necessarily the Higgs particle) in~\cite{Rosdolsky:1970, Akiba:1965}. It was found that for the Standard Model Higgs particle, bound 
states occur only if $m_{H}> 1.3$ TeV. Since this is an order of magnitude higher than the observed mass of the Higgs particle, we have to conclude that the Higgs does not form a bound state with itself. Furthermore, even for such heavy 
Higgs bosons the binding was weak.

Another way to treat the problem of relativistic two-particle bound states in SM involves the variational method within the Hamiltonian formalism of quantum field theory~\cite{Leo:1992, Xiao:1990, Darewych:1992}. This method can be extended to accommodate three-particle systems~\cite{Leo:1993, Darewych:1993, Berseth:1991}. In principle, the variational method does not depend on the coupling strength, in contrast with perturbation theory which becomes increasingly suspect as the coupling becomes stronger. This fact is relevant since the Higgs self-coupling approaches a strong regime as the Higgs boson mass becomes large, rendering perturbation theory results questionable. 
It has been estimated that the Higgs boson mass at which perturbation theory ceases  to act is approximately 
700 GeV~\cite{Gunion:1989we, Lee:1977} whereupon the theory behaves like a strongly coupled one. Gunion and others 
treated the possibility of heavier Higgs to be discovered at the LHC and hence applied the variational method rather
than the perturbative method.

Leo and Darewych in their work~\cite{Leo:1992, Xiao:1990, Darewych:1992} found that two-Higgs bound states 
which they called ``Higgsonium" would appear only for rather obese minimal Standard Model Higgs particles 
with mass $m_{H}>$ 894 GeV. This was quite similar to the 810 GeV estimate~\cite{Grifols:1991} which was 
obtained by using a phenomenological Yukawa potential to describe the Higgs-Higgs interaction.

In the approach of Bethe and Salpeter, which we mention for the sake of completeness~\cite{Bethe:1951}, 
the relativistic S-matrix formalism of Feynman was applied to the bound-state 
problem for two interacting Fermi-Dirac particles. The bound state was described by a wave function depending 
on separate times for each of the two particles forming the bound state. Integral equations 
for this wave function were derived with kernels in the form of an expansion in powers of $g^{2}$, the 
dimensionless coupling constant for the interaction. Each term in these expansions gave Lorentz-invariant 
equations. The validity and physical significance of these equations were discussed. In extreme 
non-relativistic approximation and to lowest order in $g^{2}$ they reduced to the appropriate 
Schr\"{o}dinger equation.

%
\section{Formalism of the two Higgs doublet  model}

The Lagrangian density of the two Higgs doublet model containing two Higgs doublets $\phi_{1}$ and $\phi_{2}$ of hypercharge
$\frac{1}{2}$ is given by
\begin{equation}
{\cal L}_{\phi_{1,2}}^{4}= (D^{\mu}\phi_{1})^{\dagger}(D_{\mu}\phi_{1})
+(D^{\mu}\phi_{2})^{\dagger}(D_{\mu}\phi_{2})
-V(\phi_{1,2})+ h.c.\,,\label{Lagrangian.L4}
\end{equation}
where the covariant derivative is 
\begin{equation}
D_{\mu}=\partial_{\mu}-ig_{1}B_{\mu}-ig_{2}\frac{\sigma^{I}}{2}
W_{\mu}^{I}\,,
\end{equation}
$\sigma^{I}$ are the Pauli matrices and $W_{\mu}^{I}$ and $B_{\mu}$ are SU(2) and U(1) gauge boson operators. We will work with the  scalar potential~\cite{Lee:1973iz, Gunion:1989we}
\begin{eqnarray}
V(\phi_{1,2}) &=&
\lambda_{1}\left(|\phi_{1}|^2 
- \frac{v_{1}^{2}}{2}\right)^{2} + 
\lambda_{2}\left(|\phi_{2}|^2 
- \frac{v_{2}^{2}}{2}\right)^{2}     
 \nonumber \\
 & &\quad 
+\lambda_3\left(|\phi_1|^2 + |\phi_2|^2
-\frac{v_{1}^{2} +  v_{2}^{2}}{2}\right)^{2}  
\nonumber \\ 
&& \quad
+\lambda_{4}\left(|\phi_{1}|^2 |\phi_{2}|^2 
- |\phi_{1}^{\dagger}\phi_{2}|^2\right) 
 \nonumber \\
 & &\quad 
 + \lambda_5\left|\phi_1^\dagger\phi_2 -  \frac{v_1v_2}{2}\right|^2\,,
\label{2HDM.potential}
\end{eqnarray}
where the $\lambda_i$ are real parameters. To avoid flavor-changing neutral currents 
(FCNCs)~\cite{Glashow:1976nt, Paschos:1976ay}, we impose an additional U(1) symmetry. The potential $V(\phi_{1,2})$ is
invariant under the symmetry $\phi_1 \to e^{i\theta}\phi_1\,, \phi_2 \to \phi_2\,,$
except for a soft breaking term $\lambda_5 v_1v_2 \Re(\phi_{1}^{\dagger} \phi_{2})\,.$ Additional dimension four terms,
including one allowed by a softly broken $Z_2$ symmetry~\cite{Gunion:1992hs}
are also set to zero by this U(1) symmetry. One such term was $\lambda_{6}(\frac{1}{2i}(\phi_{1}^{\dagger}\phi_{2}-\phi_{2}^{\dagger}\phi_{1}))^{2}$\,.

We consider the new physics scale at ${\cal M}\sim 1$ TeV. Now the 2HDM Lagrangian density is supplemented with higher dimension operators. The effective Lagrangian density of this extended 2HDM can be written as
\begin{equation}
{\cal L}_{\phi_{1,2}}={\cal L}_{\phi_{1,2}}^{4}+\frac{{\cal L}_{\phi_{1,2}}^{6}}{{\cal M}^{2}}+{\cal O} \left(\frac{v_{1}^{4}}{{\cal M}^{4}}\right)+{\cal O} \left(\frac{v_{2}^{4}}{{\cal M} ^{4}}\right)\,,\label{Lagrangian.Total}
\end{equation}
where the dimension six operators that preserve the symmetries of the 2HDM (here U(1) symmetry) and
custodial $SU(2)_{C}$~\cite{Solberg:2018} in the Higgs sector are given by
\begin{eqnarray}
{\cal L}_{\phi_{1,2}}^{6} &\rightarrow&
{\cal L}_{C}^{6}=C_{\phi_{1}}^{1}\partial^{\mu}(\phi_{1}^{\dagger}\phi_{1})
\partial_{\mu}(\phi_{1}^{\dagger}\phi_{1})+ C_{\phi_{1}}^{2}(\phi_{1}^{\dagger}\phi_{1})
(D_{\mu}\phi_{1})^{\dagger}(D^{\mu}\phi_{1})-\frac{\lambda_{7}}{3!}(\phi_{1}^{\dagger}\phi_{1})^{3}\nonumber \\
 & &\quad 
 + C_{\phi_{2}}^{3}\partial^{\mu}(\phi_{2}^{\dagger}\phi_{2})
\partial_{\mu}(\phi_{2}^{\dagger}\phi_{2})+ C_{\phi_{2}}^{4}(\phi_{2}^{\dagger}\phi_{2})
(D_{\mu}\phi_{2})^{\dagger}(D^{\mu}\phi_{2})-\frac{\lambda_{8}}{3!}(\phi_{2}^{\dagger}\phi_{2})^{3}\nonumber \\
& &\quad
 + C_{\phi_{1,2}}^{5}\partial^{\mu}(\phi_{1}^{\dagger}\phi_{2})^{\dagger}
\partial_{\mu}(\phi_{1}^{\dagger}\phi_{2})-\frac{1}{2}\phi_{1}^{\dagger}\phi_{2}\phi_{2}^{\dagger}\phi_{1}(\lambda_{9}\phi_{1}^{\dagger}\phi_{1}+\lambda_{9}^{\prime}\phi_{2}^{\dagger}\phi_{2})\,. 
\label{Lagrangian.L6}
\end{eqnarray}
Appendix~(\ref{Appendix A}) discusses how the fields transform under custodial symmetry.

We expand the scalar fields about their vacuum expectation values $v_{1}$ and $v_{2}$\,, 
\begin{align}
\phi_{1}(x)&= \frac{U_{1}(x)}{\sqrt{2}}
\left(
\begin{array}{c}
0\\
v_{1}+h(x)
\end{array}
\right)\,, \\
\phi_{2}(x)&= \frac{U_{2}(x)}{\sqrt{2}}
\left(
\begin{array}{c}
0\\
v_{2}+H(x)
\end{array}
\right)\,.
\end{align}
Here $h$ and $H$ are the CP-even Higgs fields, with $\langle h(x)\rangle = 0$ and 
$\langle H(x)\rangle = 0$\,, and $U_{i}(x)=e^{i\xi_{i}^{a}(x)\sigma_{a}/v_{i}}$, $i=1,2$\,. 
The six fields $\xi_{i}^{a}$ include the three Goldstone 
bosons which get eaten by the $W^{\pm}$ and Z bosons to make them massive, while the other 
three combine to become
the charged Higgs bosons and the CP-odd Higgs boson. The vevs $v_1$ and $v_2$\,, and therefore 
$v = \sqrt{v_1^2 + v_2^2}\,,$ 
will be taken to be small compared to ${\cal M}\,,$ the scale of new physics since $v$ = 246 GeV $\ll$  1 TeV. We note that 
degrees of freedom are easier to identify when the doublets are written in terms of real and 
imaginary parts of the complex scalar fields, but for our calculations in this paper it is
more convenient to work in the unitarity gauge, in which the gauge transformation has been 
used to remove the Goldstone bosons from the Lagrangian. The charged scalars and the CP-odd 
scalar will still remain in the Lagrangian, but we can neglect their contribution for the 
bound state calculations. 

In order to normalize the kinetic term to have a coefficient of $\frac{1}{2}\,,$ we redefine the fields as
\begin{align}
h &\rightarrow \frac{h^{\prime}}{(1+2C_{h}^{K})^{1/2}}\label{h_redef}\\
H &\rightarrow \frac{H^{\prime}}{(1+2C_{H}^{K})^{1/2}}\,,\label{H_redef}
\end{align}
where
\begin{align}
C_{h}^{K} &= (v_{1}^{2}/{\cal M}^{2})(C_{\phi_{1}}^{1}
+\frac{1}{4}C_{\phi_{1}}^{2})\,, \label{ChK_redef} \\
C_{H}^{K} &= (v_{2}^{2}/{\cal M}^{2})(C_{\phi_{2}}^{3}+
\frac{1}{4}C_{\phi_{2}}^{4})\,.\label{CHK_redef}
\end{align}

We write the potential in terms of the rescaled fields, focusing on the self couplings of the CP neutral Higgs fields. We call this potential $V_{\mathit{eff}}$. Though we will be discussing the possibility of bound state formation of the heavy CP-even Higgs field, but still for the sake of completeness we will write the self couplings of the light CP-even Higgs field too. In terms of the rescaled fields, the terms in the effective potential which are of interest to us can be written as
\begin{eqnarray}
V_{\mathit{eff}}(h^{\prime},H^{\prime})& \supset & \frac{1}{2}m_{h}^{2}h^{\prime 2}+\frac{1}{2}m_{H}^{2}H^{\prime 2}
+ v_{1}\frac{\lambda_{10}^{\mathit{eff}}}{3!}h^{\prime 3}+ \frac{\lambda_{11}^{\mathit{eff}}}{4!}h^{\prime 4}+
+ v_{2}\frac{\lambda_{12}^{\mathit{eff}}}{3!}H^{\prime 3}+ \frac{\lambda_{13}^{\mathit{eff}}}{4!}H^{\prime 4}\nonumber \\
& & \quad 
+\frac{\lambda_{14}^{\mathit{eff}}}{2!}v_{2}h^{\prime}h^{\prime}H^{\prime}+\frac{\lambda_{15}^{\mathit{eff}}}{2!}v_{1}H^{\prime}H^{\prime}h^{\prime}+\frac{\lambda_{16}^{\mathit{eff}}}{2!2!}h^{\prime}h^{\prime}H^{\prime}H^{\prime}\,.\label{B}
\end{eqnarray}
The mass terms and the coupling constants are related to the original $\lambda_i$. Since we are interested in the bound state formation of $H$ therefore we write down the cubic and quartic self couplings and also the mass of $H$ in terms of the original $\lambda_{i}$ and evaluate their relative strengths.
\begin{eqnarray}
m_{H}^{2} &=& (1-2C_{H}^{K})(2v_{2}^{2}(\lambda_{2}+\lambda_{3})+\frac{\lambda_{5}}{2}v_{1}^{2})+ \frac{5\lambda_{8}}{8}\frac{v_{2}^{4}}{{\cal M}^{2}}+\frac{\lambda_{9}}{8}\frac{v_{1}^{4}}{{\cal M}^{2}}+\frac{3\lambda_{9}^{\prime}}{4}\frac{v_{1}^{2}v_{2}^{2}}{{\cal M}^{2}}\notag\\
& &+{\cal O}(\frac{v^{4}}{{\cal M}^{4}}) \label{mH} \\
\lambda_{12}^{\mathit{\mathit{eff}}}&=&6(\lambda_{2}+\lambda_{3})(1-3C_{H}^{K})+ \frac{5 \lambda_{8}}{2}
\frac{v_{2}^{2}}{{\cal M}^{2}}+ \frac{3\lambda_{9}^{\prime}}{2}\frac{v_{1}^{2}}{{\cal M}^{2}}
+{\cal O}(\frac{v^{4}}{{\cal M}^{4}}) \label{lam12}\\
\lambda_{13}^{\mathit{eff}}&=&6(\lambda_{2}+\lambda_{3})(1-4C_{H}^{K})+ \frac{15 \lambda_{8}}{2}
\frac{v_{2}^{2}}{{\cal M}^{2}}+ \frac{3\lambda_{9}^{\prime}}{2}\frac{v_{1}^{2}}{{\cal M}^{2}}
+{\cal O}(\frac{v^{4}}{{\cal M}^{4}}) \label{lam13} \\
\lambda_{14}^{\mathit{eff}}&=&(2\lambda_{3}+\lambda_{5})(1-2C_{h}^{K})(1-C_{H}^{K})+ \frac{3 \lambda_{9}}{2}
\frac{v_{1}^{2}}{{\cal M}^{2}}+ \frac{\lambda_{9}^{\prime}}{2}\frac{v_{2}^{2}}{{\cal M}^{2}}
+{\cal O}(\frac{v^{4}}{{\cal M}^{4}}) \label{lam14} \\
\lambda_{15}^{\mathit{eff}}&=&(2\lambda_{3}+\lambda_{5})(1-2C_{H}^{K})(1-C_{h}^{K})+ \frac{ \lambda_{9}}{2}
\frac{v_{1}^{2}}{{\cal M}^{2}}+ \frac{3\lambda_{9}^{\prime}}{2}\frac{v_{2}^{2}}{{\cal M}^{2}}
+{\cal O}(\frac{v^{4}}{{\cal M}^{4}}) \label{lam15}\\
\lambda_{16}^{\mathit{eff}}&=&(2\lambda_{3}+\lambda_{5})(1-2C_{h}^{K})(1-2C_{H}^{K})+ \frac{3 \lambda_{9}}{2}
\frac{v_{1}^{2}}{{\cal M}^{2}}+ \frac{3\lambda_{9}^{\prime}}{2}\frac{v_{2}^{2}}{{\cal M}^{2}}+{\cal O}(\frac{v^{4}}{{\cal M}^{4}})\label{lam16}
\end{eqnarray}

The mixed cubic and quartic Higgs boson couplings corresponding to the 
coupling constants $\lambda_{14}^{\mathit{eff}}$, $\lambda_{15}^{\mathit{eff}}$ and
$\lambda_{16}^{\mathit{eff}}$ are needed for the $h-H$ bound state, but since the calculations for that are much 
more involved, we will leave the study of that bound state for another occasion. As regards the fermionic operators, since they are not needed in the theory for bound state formation we do not explicitly show them here.


We should mention here that for a single Higgs particle, the effective field theory may also be written 
as a nonlinear realization analogous 
to the $\sigma$ and $\pi$ fields of QCD as described by a chiral Lagrangian~\cite{Grinstein:2007iv}. It is 
not obvious to us how to write a nonlinear realization involving neutral and charged Higgs fields along with 
the necessary Goldstone bosons, nor is it clear whether that will help in looking for bound states. So we 
will stick to the linear realization.

\section{Effective couplings at low energy}\label{L2HET}

When we discuss physics at the low energy scale, the heavier momentum modes need to be integrated out from the theory. As the top Yukawa coupling is fairly large compared to the other fermions, we need to estimate the effects of the top quark on the possibility of a bound state of $H$. When the top quark mass is much heavier than the Higgs mass, the effect of integrating out the top quark shows up in the modified coupling constants and masses of the Higgs particle.

We will use this approximation even when the Higgs is slightly heavier than the top. 
As mentioned in~\cite{Grinstein:2007iv}, for the case of a single Higgs field, the 
approximation is known to work better than one would expect when $m_{h}<2m_{t}$. 
This is because of the absence of any non-analytic dependence on the mass since the Higgs is the pseudo-goldstone boson of spontaneously broken scale invariance~\cite{Grinstein:1988,Dawson:1992, Chivukula:1989}. However, when the mass of the Higgs particle is more than $2m_{t}$, we cannot integrate out the top quark. We will work in the \textit{Alignment limit} and thus we must set $m_{h}=125$ GeV. The remaining heavier CP-even Higgs can have any mass above 125 GeV restricted by constraints coming from perturbative unitarity and stability. In~\cite{Biswas:2014uba} its mass was further restricted 
by use of Naturalness conditions, and the bounds were found to be 450 GeV $\lesssim m_{H} \lesssim $ 
620 GeV for $\tan\beta =$5. It is worth mentioning here that though these limits on $m_{H}$ are for Type - II 2HDM but the other types of 2HDMs also exhibit the mass ranges for $H$ in the close vicinity of these limits. Moreover when these mass ranges were evaluated the most recent value of $\rho$ - parameter was used~\cite{Yao:2006}. In this paper we will usually work with these limits, but also consider the possibility that the heavier Higgs has a mass smaller than $2m_{t}$. We will not consider the situation where the heavier CP-even Higgs is identified with the Standard Model Higgs (Reverse Alignment limit), for reasons discussed in~\cite{Biswas:2015zgk}.

Let us also mention here our choice for the parameters used in the calculations. There is 
no bound on the value of $\tan\beta$, which is perhaps 
the most important parameter in the 2HDMs, except that it should be larger than unity. 
This is based on constraints coming from $Z\rightarrow b\bar{b}$, $B_{q}\bar{B}_{q}$
mixing~\cite{Benbrik:2009}, muon $g-2$ in lepton specific 2HDM~\cite{Wan:2009} or using 
$b\rightarrow s\gamma$ in type I and flipped models~\cite{Park:2006}. Thus we take $\tan\beta=$5 which is a reasonable choice, $v=246$ GeV, $m_{t}=174$ GeV and the new physics scale ${\cal M}$ to be 1TeV. We broadly categorize the heavier Higgs boson mass as $m_{H}<2m_{t}$ and $m_{H}>2m_{t}$. For $m_{H}<2m_{t}$ the top quark is integrated out while for $m_{H}>2m_{t}$, the effect of top quark is retained in the theory.

\subsection*{Integrating out the top quark: $m_{H}\lesssim 2m_{t}$}

\begin{figure}[htbp]
\includegraphics[height=0.25\columnwidth, width = 1.0\columnwidth]{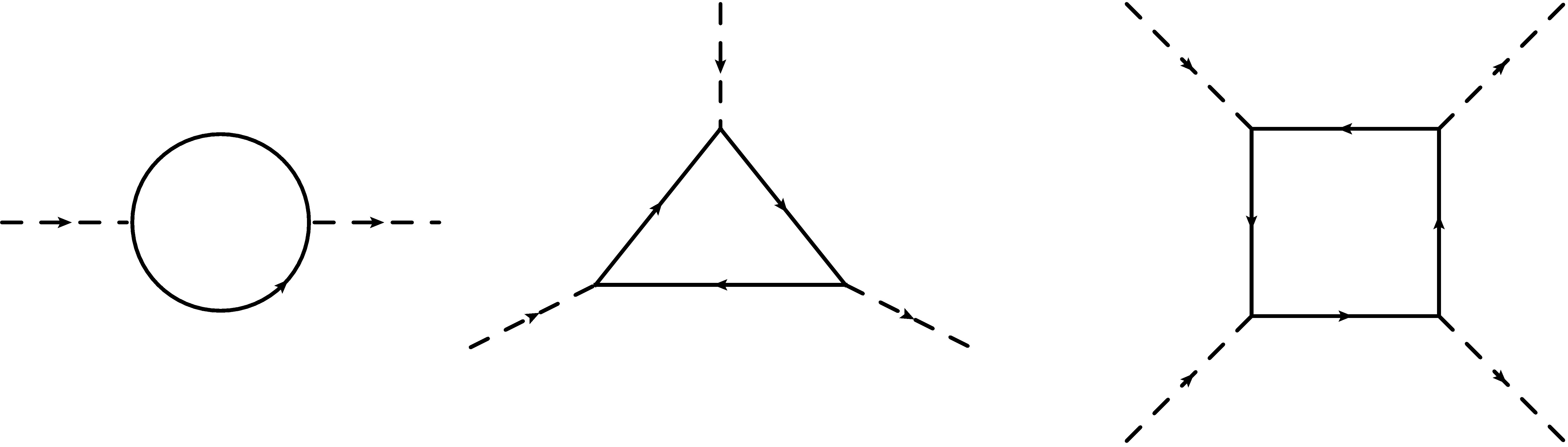} 
\caption{$t$-quark loop corrections to Higgs self couplings}
\label{Top-Higgs.fig}
\end{figure}

The top mass term and couplings to the Higgs bosons are given by
\begin{equation}
{\cal L}_{Y}=-\dfrac{m_{t}}{v}\xi_{H}^{t}\overline{t}tH\,.
\end{equation}
where $\xi$ stands for the Yukawa coupling of $H$ with the fermion indicated in the superscript.
The values of $\xi$ for up-type and down-type quarks are displayed in Table~\ref{Yukawa.Table}.
 \begin{table}[tbhp]
\begin{tabular}{||c|c|c||}
\hline
2HDMs & $\; \xi_{H}^{u}$ &$\; \xi_{H}^{d}$ \\
\hline
Type I &$\;\frac{\sin\alpha}{\sin\beta} $
&$\; \frac{\sin\alpha}{\sin\beta} $\\
\hline 
Type II &$\;\frac{\sin\alpha}{\sin\beta} $
&$\; \frac{\cos\alpha}{\cos\beta} $\\
\hline 
Lepton Specific &$\;\frac{\sin\alpha}{\sin\beta} $
&$\; \frac{\sin\alpha}{\sin\beta} $\\
\hline 
Flipped &$\;\frac{\sin\alpha}{\sin\beta} $
&$\; \frac{\cos\alpha}{\cos\beta} $\\
\hline
\end{tabular}
\caption{Yukawa couplings for the different 2HDMs}
\label{Yukawa.Table}
\end{table}
Fig.~\ref{Top-Higgs.fig} shows the Feynman graphs that contribute to modifications of the Higgs self-couplings. The solid line denotes a top quark, the external dashed lines denote the heavy CP-even neutral Higgs boson.
Calculations are performed up to the lowest order in $p^{2} /m_{t}^{2}$\,. 
These corrections further modify the 
effective potential of the Higgs scalar field $H$. For example, the 1-loop correction to 
the four point function of $H$ requires the four point amplitude with the top quark 
circulating in the loop, which has the form
\begin{equation}
i{\cal A}_{4}(s,t,u)=-6N_{c}\left(\dfrac{m_{t}}{v}\xi_{H}^{t}\right)^{4}\int\dfrac{d^{d}k}
{(2\pi)^{d}}\Tr\left[\dfrac{(\slashed{k}+m_{t})(\slashed{k}+\slashed{a}+m_{t})(\slashed{k}+
	\slashed{b}+m_{t})(\slashed{k}+\slashed{c}+m_{t})}{(k^{2}-m_{t}^{2})((k+a)^{2}-m_{t}^{2})
	((k+b)^{2}-m_{t}^{2})((k+c)^{2}-m_{t}^{2})}\right]\,.
\end{equation}
%
%
For leading order in $p^{2} /m_{t}^{2}\rightarrow 0$, the amplitude is given by
\begin{align}
i{\cal A}_{4}^{0}(s,t,u)&= -24N_{c}\left(\frac{m_{t}}{v}\xi_{H}^{t}\right)^{4}
\int\dfrac{d^{d}k}{(2\pi)^{d}}\frac{(m_{t}^{4}+6k^{2}m_{t}^{2}
	+k^{4})}{(k^{2}-m_{t}^{2})^{4}} \notag \\
&=-\frac{iN_{c}}{16\pi^{2}}\left(\frac{m_{t}}{v}\xi_{H}^{t}\right)^{4}
\left(\frac{24}{\epsilon}-64+24\log\left(\frac{\mu^{2}}{m_{t}^{2}}\right)\right)\,.
\end{align}
The leading order term gives a factor of\, $-\dfrac{4N_{c}}{\pi^{2}}\left(\dfrac{m_{t}}{v}\xi_{H}^{t}\right)^{4}$. 
Similar calculations have been done for the other effective couplings and mass terms. Thus the expressions for the effective 
couplings and mass terms given by Eqs.~(\ref{mH})~--~(\ref{lam16}) get modified as 
\begin{eqnarray}
m_{H}^{2} &=& (1-2C_{H}^{K})(2v_{2}^{2}(\lambda_{2}+\lambda_{3})+\frac{\lambda_{5}}{2}v_{1}^{2})+ \frac{5\lambda_{8}}{8}\frac{v_{2}^{4}}{{\cal M}^{2}}+\frac{\lambda_{9}}{8}\frac{v_{1}^{4}}{{\cal M}^{2}}+\frac{3\lambda_{9}^{\prime}}{4}\frac{v_{1}^{2}v_{2}^{2}}{{\cal M}^{2}}+\frac{N_{c}}{4\pi^{2}}\frac{m_{t}^{4}}{v^{2}}(\xi_{H}^{t})^{2}\notag\\
& & +{\cal O}(\frac{v^{4}}{{\cal M}^{4}}) 
\label{mH.1}\\
\lambda_{12}^{\mathit{eff}}&=&6(\lambda_{2}+\lambda_{3})(1-3C_{H}^{K})+ \frac{5 \lambda_{8}}{2}
\frac{v_{2}^{2}}{{\cal M}^{2}}+ \frac{3\lambda_{9}^{\prime}}{2}\frac{v_{1}^{2}}{{\cal M}^{2}}-\frac{N_{c}}{\pi^{2}}\frac{m_{t}^{4}}{v^{4}}(\xi_{H}^{t})^{3}+{\cal O}(\frac{v^{4}}{{\cal M}^{4}})\label{lam12.1}\\
\lambda_{13}^{\mathit{eff}}&=&6(\lambda_{2}+\lambda_{3})(1-4C_{H}^{K})+ \frac{15 \lambda_{8}}{2}
\frac{v_{2}^{2}}{{\cal M}^{2}}+ \frac{3\lambda_{9}^{\prime}}{2}\frac{v_{1}^{2}}{{\cal M}^{2}}-\frac{4N_{c}}{\pi^{2}}\frac{m_{t}^{4}}{v^{4}}(\xi_{H}^{t})^{4}+{\cal O}(\frac{v^{4}}{{\cal M}^{4}})\label{lam13.1}\\
\lambda_{14}^{\mathit{eff}}&=&(2\lambda_{3}+\lambda_{5})(1-2C_{h}^{K})(1-C_{H}^{K})+\frac{3 \lambda_{9}}{2}\frac{v_{1}^{2}}{{\cal M}^{2}}+ \frac{\lambda_{9}^{\prime}}{2}\frac{v_{2}^{2}}{{\cal M}^{2}}-\frac{N_{c}}{6\pi^{2}}\frac{m_{t}^{4}}{v^{4}}(\xi_{h}^{t})^{2}(\xi_{H}^{t})\notag\\
& &+{\cal O}(\frac{v^{4}}{{\cal M}^{4}})\label{lam14.1}\\
\lambda_{15}^{\mathit{eff}}&=&(2\lambda_{3}+\lambda_{5})(1-2C_{H}^{K})(1-C_{h}^{K})+ \frac{ \lambda_{9}}{2}\frac{v_{1}^{2}}{{\cal M}^{2}}+ \frac{3\lambda_{9}^{\prime}}{2}\frac{v_{2}^{2}}{{\cal M}^{2}}-\frac{N_{c}}{6\pi^{2}}\frac{m_{t}^{4}}{v^{4}}(\xi_{h}^{t})(\xi_{H}^{t})^{2}\notag\\
& &+{\cal O}(\frac{v^{4}}{{\cal M}^{4}})\label{lam15.1}\\
\lambda_{16}^{\mathit{eff}}&=&(2\lambda_{3}+\lambda_{5})(1-2C_{h}^{K})(1-2C_{H}^{K})+ \frac{3 \lambda_{9}}{2}\frac{v_{1}^{2}}{{\cal M}^{2}}+ \frac{3\lambda_{9}^{\prime}}{2}\frac{v_{2}^{2}}{{\cal M}^{2}}-\frac{4N_{c}}{24\pi^{2}}\frac{m_{t}^{4}}{v^{4}}(\xi_{h}^{t})^{2}(\xi_{H}^{t})^{2}\notag\\
& &+{\cal O}(\frac{v^{4}}{{\cal M}^{4}})\label{lam16.1}
\end{eqnarray}

where $N_{c}$ stands for the three colors of the top quark. Contributions of other quarks 
as the loop particle have been ignored here because of the very large difference in the 
masses of the top quark and the other quarks. These are the effective low energy couplings which have been obtained by integrating out the heavier momentum modes. They will be used in the next section to study the possibility of formation of H-H bound state. 

\section{Phenomenology of bound state formation}
Due to the D = 6 operators the three and four point contact interactions and the Higgs mass $m_{H}$ gain corrections in the effective 
potential. Eliminating the self-couplings $\lambda_{1}\,, \lambda_{2}\,$ and $\lambda_{3}$ in favour 
of the Higgs mass $m_{H}$\,, we can write for the effective cubic and quartic Higgs-self couplings,
\begin{eqnarray}
\lambda_{12}^{\mathit{eff}} &=& 3(1-C_{H}^{K})\frac{m_{H}^{2}}{v_{2}^{2}}-\frac{3\lambda_{5}}{2}\frac{v_{1}^{2}}{v_{2}^{2}}(1-3C_{H}^{K})+\frac{5\lambda_{8}}{8}\frac{v_{2}^{2}}{{\cal M}^{2}}-\frac{3\lambda_{9}}{8}\frac{v_{1}^{4}}{{\cal M}^{2}v_{2}^{2}}(1-C_{H}^{K}) \notag \\ 
& & -\frac{3\lambda_{9}^{\prime}}{4}\frac{v_{1}^{2}}{{\cal M}^{2}}
-(7-3C_{H}^{K})\frac{N_{c}}{4\pi^{2}}\left(\frac{m_{t}^{4}}{v^{4}}\right)\left(\xi^{t}_{H}\right)^{3}\,,\label{lam12_A}\\
\lambda_{13}^{\mathit{eff}} &=& 3(1-2C_{H}^{K})\frac{m_{H}^{2}}{v_{2}^{2}} 
-\frac{3\lambda_{5}}{2}\frac{v_{1}^{2}}{v_{2}^{2}}(1-4C_{H}^{K})+\frac{45\lambda_{8}}{8}\frac{v_{2}^{2}}{{\cal M}^{2}}-\frac{3\lambda_{9}}{8}\frac{v_{1}^{4}}{{\cal M}^{2}v_{2}^{2}}(1-2C_{H}^{K}) \notag\\
& & -\frac{3\lambda_{9}^{\prime}}{4}\frac{v_{1}^{2}}{{\cal M}^{2}}-(19-6C_{H}^{K})\frac{N_{c}}{4\pi^{2}}\left(\frac{m_{t}^{4}}{v^{4}}\right)\left(\xi^{t}_{H}\right)^{4}\,.\label{lam13_A}
\end{eqnarray}

We consider the non-relativistic Schr\" odinger equation to gain some idea about the bound state formation. It reads
\begin{equation}
[-\bigtriangledown_{r}^{2}+V(r)-E]\psi(r)=0\,.
\label{Schrodinger.eq}
\end{equation}
The above potential has contributions from a Yukawa exchange and a contact interaction,
\begin{equation}
V(r)=-\frac{g^{2}}{4\pi}\frac{e^{-mr}}{r}+\kappa\delta^{3}(r)\,,
\label{Schrodinger.eq.potential}
\end{equation}
where $g$ denotes the Yukawa exchange coupling constant and $\kappa$ denotes the contact interaction coupling constant for the two CP even neutral Higgs bosons.
As a non-relativistic approximation of the Higgs bosons self interactions, $g$ corresponds to 3 point coupling and $\kappa$ corresponds to 4 point coupling.
Thus from the potential in Eq.~(\ref{Schrodinger.eq.potential}) we can conclude that the 
attractive contact interaction and the repulsive contact interaction are governed by the 
cubic and quartic couplings respectively. If the attractive interaction overpowers 
the repulsive interaction, the formation of a bound state becomes feasible. Let us consider 
the formation of the Higgs-Higgs bound states by evaluating the relative strengths of the 
cubic and quartic self couplings.

\subsubsection{H-H bound state}
\subsubsection*{Category I: $m_{h}\leq m_{H} \leq 2m_{t}$}
When the cancellation of quadratic divergences is used as a criterion of restriction, 
$m_H$ turns out to be heavier than $2m_t$~\cite{Biswas:2014uba}\,. However, the idea 
of an $H-H$ bound state 
is not restricted by the consideration of naturalness, so it is worthwhile to check on
the possibility of bound state formation even when $m_{h}\leq m_{H} \leq 2m_{t}$\,.
In this case we integrate out the top quark as discussed earlier. Then, using 
Eq.~(\ref{CHK_redef}) in Eqs.~(\ref{lam12_A}) and (\ref{lam13_A}) we can calculate
the cubic and quartic couplings of the $H$ field,
\begin{eqnarray}
\lambda_{12}^{\mathit{eff}} 
&=& \frac{3m_{H}^{2}}{v^{2}\tan^{2}\beta}(1+\tan^{2}\beta)-\frac{3m_{H}^{2}}{{\cal M}^{2}}\left(C_{\phi_{2}}^{3}+\frac{1}{4}C_{\phi_{2}}^{4}\right)-\frac{3\lambda_{5}}{2\tan^{2}\beta} +\frac{5\lambda_{8}}{8}\frac{v^{2}\tan^{2}\beta}{{\cal M}^{2}(1+\tan^{2}\beta)} \notag \\
& & - \frac{3\lambda_{9}}{8{\cal M}^{2}}\frac{v^{2}}{(1+\tan^{2}\beta)\tan^{2}\beta}
-\frac{3\lambda_{9}^{\prime}}{4}\frac{v^{2}}{{\cal M}^{2}(1+\tan^{2}\beta)} +\frac{7N_{c}}{4\pi^{2}}\left(\frac{m_{t}^{4}}{v^{4}}\right)\cot^{3}\beta\,,
\label{lam12_B}\\
\lambda_{13}^{\mathit{eff}} 
&=& \frac{3m_{H}^{2}}{v^{2}\tan^{2}\beta}(1+\tan^{2}\beta)-\frac{6m_{H}^{2}}{{\cal M}^{2}}\left(C_{\phi_{2}}^{3}+\frac{1}{4}C_{\phi_{2}}^{4}\right)-\frac{3\lambda_{5}}{2\tan^{2}\beta} +\frac{45\lambda_{8}}{8}\frac{v^{2}\tan^{2}\beta}{{\cal M}^{2}(1+\tan^{2}\beta)} \notag \\
& & - \frac{3\lambda_{9}}{8{\cal M}^{2}}\frac{v^{2}}{(1+\tan^{2}\beta)\tan^{2}\beta}
-\frac{3\lambda_{9}^{\prime}}{4}\frac{v^{2}}{{\cal M}^{2}(1+\tan^{2}\beta)}-\frac{19N_{c}}{4\pi^{2}}
\left(\frac{m_{t}^{4}}{v^{4}}\right)\cot^{4}\beta\,.
\label{lam13_B}
\end{eqnarray}
We have put $\xi_{H}^{t}\approx -\cot\beta\,,$ which is its value for all types of 
2HDMs in the alignment limit. Letting $(C_{\phi_{2}}^{3}+\frac{1}{4}C_{\phi_{2}}^{4})\sim 1$, 
for $v= 246$ GeV, $m_{t}=174$ GeV, $\tan \beta=5$ 
and keeping the $\lambda$'s well within the perturbative bounds by choosing 
$\lambda_{i}\sim 1\,,$ we have evaluated 
the strengths of the cubic and quartic couplings from Eqs.~(\ref{lam12_B}) and (\ref{lam13_B}) for $m_{H}$ = 300 GeV.
We have found that $|\lambda_{12}^{\mathit{eff}}| - |\lambda_{13}^{\mathit{eff}}|\,= -0.02\,,$
i.e. $|\lambda_{12}^{\mathit{eff}}| \approx |\lambda_{13}^{\mathit{eff}}|$ at the level 
of accuracy we are considering. We conclude that in this case of a not too heavy $H\,,$ 
an $H-H$ bound state may form, but it is also likely to have a very short lifetime.

\subsubsection*{Category II: $m_{H} \geq 2m_{t}$}
%
Naturalness arguments coupled with unitarity, perturbativity and constraints from the T-parameter lead to a heavy $H$ with a mass between 450 GeV and 620 GeV~\cite{Biswas:2014uba}\,. In this case we cannot integrate out the top quark. For the cubic and quartic couplings we find
\begin{eqnarray}
\lambda_{12}^{\mathit{eff}}&=& 3(1-C_{H}^{K})\frac{m_{H}^{2}}{v_{2}^{2}}-\frac{3\lambda_{5}}{2}\frac{v_{1}^{2}}{v_{2}^{2}}(1-3C_{H}^{K})+\frac{5\lambda_{8}}{8}\frac{v_{2}^{2}}{{\cal M}^{2}}-\frac{3\lambda_{9}}{8}\frac{v_{1}^{4}}{{\cal M}^{2}v_{2}^{2}}(1-C_{H}^{K}) 
-\frac{3\lambda_{9}^{\prime}}{4}\frac{v_{1}^{2}}{{\cal M}^{2}} \nonumber \\
&=& \frac{3 m_{H}^{2}}{v^{2}\tan^{2}\beta}(1+\tan^{2}\beta)-\frac{3m_{H}^{2}}{{\cal M}^{2}}(C_{\phi_{2}}^{3}+\frac{1}{4}C_{\phi_{2}}^{4})-\frac{3\lambda_{5}}{2\tan^{2}\beta} +\frac{5\lambda_{8}}{8}\frac{v^{2}\tan^{2}\beta}{{\cal M}^{2}(1+\tan^{2}\beta)} \nonumber\\
& &  -\frac{3\lambda_{9}}{8{\cal M}^{2}}\frac{v^{2}}{(1+\tan^{2}\beta)\tan^{2}\beta}
-\frac{3\lambda_{9}^{\prime}}{4}\frac{v^{2}}{{\cal M}^{2}(1+\tan^{2}\beta)}\,,\label{lam12_C}\\
\lambda_{13}^{\mathit{eff}}&=& 3(1-2C_{H}^{K})\frac{m_{H}^{2}}{v_{2}^{2}}-\frac{3\lambda_{5}}{2}\frac{v_{1}^{2}}{v_{2}^{2}}(1-4C_{H}^{K})+\frac{45\lambda_{8}}{8}\frac{v_{2}^{2}}{{\cal M}^{2}}-\frac{3\lambda_{9}}{8}\frac{v_{1}^{4}}{{\cal M}^{2}v_{2}^{2}}(1-2C_{H}^{K}) 
-\frac{3\lambda_{9}^{\prime}}{4}\frac{v_{1}^{2}}{{\cal M}^{2}} \nonumber \\
&=& \frac{3 m_{H}^{2}}{v^{2}\tan^{2}\beta}(1+\tan^{2}\beta)-\frac{6m_{H}^{2}}{{\cal M}^{2}}(C_{\phi_{2}}^{3}+\frac{1}{4}C_{\phi_{2}}^{4})-\frac{3\lambda_{5}}{2\tan^{2}\beta} +\frac{45\lambda_{8}}{8}\frac{v^{2}\tan^{2}\beta}{{\cal M}^{2}(1+\tan^{2}\beta)} \nonumber\\
& &  -\frac{3\lambda_{9}}{8{\cal M}^{2}}\frac{v^{2}}{(1+\tan^{2}\beta)\tan^{2}\beta}-\frac{3\lambda_{9}^{\prime}}{4}\frac{v^{2}}{{\cal M}^{2}(1+\tan^{2}\beta)}\,.\label{lam13_C}
\end{eqnarray}
Letting $(C_{\phi_{2}}^{3}+\frac{1}{4}C_{\phi_{2}}^{4})\sim 1$, for $v=246$ GeV, $m_{t}=174$ GeV, $\tan \beta=5$ 
and keeping the $\lambda$'s well within the perturbative bounds by choosing $\lambda_{i}\sim 1\,,$ the strengths of the cubic and quartic couplings were evaluated from Eqs.~(\ref{lam12_C}) and (\ref{lam13_C}) for $m_{H}$ = 450 GeV, 500 GeV and 620 GeV. The values 450 GeV and 620 GeV are the lower and upper limits for $m_{H}$ as found in~\cite{Biswas:2014uba} and we have also considered the intermediate value of 500 GeV. We found that $|\lambda_{12}^{\mathit{eff}}| - |\lambda_{13}^{\mathit{eff}}|\,=0.317\;, 0.459$ and $0.87$ respectively for $m_{H}$ = 450 GeV, 500 GeV and 620 GeV.
Thus formation of $H-H$ bound state is likely for obese Higgs bosons and the likelihood increases as the Higgs becomes more massive.

\subsection{HIGGSIUM: Production and Decay}

The formation time of the $H-H$ bound state can be approximated by $\tau_{f}^{H}\sim \frac{4R_{0}^{H}}{u_{H}}$ where $R_{0}^{H}$ is the characteristic radius of the $H-H$ bound state and $u_{H}$ is the relative velocity of the two heavier Higgs bosons. 
This is roughly the period of oscillation for $s$-wave states~\cite{Strassler:1991}.

For a non-relativistic bound state we can approximate the relative momenta of $H$ by $p_{H}\sim m_{H}u_{H}$ so that
\begin{align}
\tau_{f}^{H}&\sim \frac{4R_{0}^{H}}{u_{H}}\notag \\
 &\sim \frac{4}{m_{H}u_{H}^{2}}\,.
\end{align}

The predominant decay channel/s for $m_{h}\leq m_{H} \leq 2m_{t}$ is $H \rightarrow b \overline{b}$ and for $m_{H}> 2m_{t}$ is $H \rightarrow b \overline{b}$ and $H \rightarrow t \overline{t}$\,. We take these decays as dictating the decay rate of Higgsium. Below we calculate the decay width neglecting the effects of new physics operators.

\subsubsection*{Case I: $m_{h}<m_{H}< 2m_{t}$}
%
Again we consider a not too heavy $H$ for the sake of completeness. Neglecting the effect of the new physics operators, 
we have calculated the $H \rightarrow b \overline{b}$ decay width and 
decay time to be 
\begin{equation}
\Gamma_{b}^{H} = \frac{m_{H}(\xi_{H}^{b})^{2}}{8\pi}(1-4\frac{m_{b}^{2}}{m_{H}^{2}})^{3/2}
\approx \frac{m_{H}(\xi_{H}^{b})^{2}}{8\pi}\,, \label{gamma_bH}
\end{equation}
and
\begin{equation}
\tau_{b}^{H} = \frac{1}{\Gamma_{b}^{H}} 
= \frac{8\pi}{m_{H}(\xi_{H}^{b})^{2}}\,.\label{tau_bH}
\end{equation}
Since $m_{b}^{2}\ll m_{H}^{2}$ we have neglected $4\frac{m_{b}^{2}}{m_{H}^{2}}$ 
in comparison to 1. Let us find an estimate of $\tau_{b}^{H}$ for the given range 
of $m_{H}$. For a reasonable choice of $\tan \beta = 5$, $\tau_{b}^{H}$ for 
$m_{H}=$ 130 and 350 GeV for various types of 2HDMs in the alignment limit have 
been tabulated below in Table~\ref{Decay time.Table.1}. Note that we use the conversion 1 GeV$^{-1} = 6.58 \times 10^{-25}$ sec to find the decay time in seconds.
 \begin{table}[tbhp]
\begin{tabular}{||c|c|c|c|c||}
\hline
2HDMs & $\xi_{H}^{b}\; $
&$\;  \tau_{b}^{H}$ &$\; \tau_{b}^{H}$ ($m_{H}=$ 130 GeV) &$\; \tau_{b}^{H}$ ($m_{H}=$ 350 GeV) \\
&&& in secs & in secs\\
\hline
Type I &$\;-\cot\beta $ 
&$\; \frac{8\pi}{m_{H}\cot^{2}\beta} $  &3.18$\;\times \mathrm{10^{-24}}$
&1.18$\;\times \mathrm{10^{-24}}$\\
\hline 
Type II &$\;\tan \beta $ 
&$\; \frac{8\pi}{m_{H}\tan^{2}\beta} $  &5.09$\;\times \mathrm{10^{-27}}$
&1.89$\;\times \mathrm{10^{-27}}$\\
\hline 
Lepton Specific &$\;-\cot\beta $ 
&$\; \frac{8\pi}{m_{H}\cot^{2}\beta} $  &3.18$\;\times \mathrm{10^{-24}}$
&1.18$\;\times \mathrm{10^{-24}}$\\
\hline 
Flipped &$\;\tan\beta $ 
&$\; \frac{8\pi}{m_{H}\tan^{2}\beta} $  &5.09$\;\times \mathrm{10^{-27}}$
&1.89$\;\times \mathrm{10^{-27}}$\\
\hline
\end{tabular}
\caption{\begin{small}
		Decay time (in seconds) of $H$ when $m_{h}<m_{H}< 2m_{t}$ for the different 2HDMs.
\end{small} 
}
\label{Decay time.Table.1}
\end{table}

For the $H-H$ bound state to be formed, the formation time of the $H-H$ bound state must be smaller than the decay time of $H$. 
In other words, 
\begin{align}
\tau_{f}^{H} &< \tau_{b}^{H}\notag \\
\Rightarrow \frac{4}{m_{H}u_{H}^{2}} &< \frac{8\pi}{m_{H}(\xi_{H}^{b})^{2}}\notag \\
 \Rightarrow\qquad u_{H} &> \frac{\xi_{H}^{b}}{\sqrt{2\pi}}\,.
\label{relative_velocity_limit.1}
 \end{align}
Now we proceed in two ways. First we fix a value of $\tan\beta$ consistent with 
observations~\cite{Benbrik:2009, Wan:2009, Park:2006} and find the range of $u_{H}$ for
which a bound state may form.
Next we fix a non-relativistic value of $u_{H}$ and find the range for $\tan\beta$.

Let us fix $\tan\beta = 5$  and use Eq.~(\ref{relative_velocity_limit.1}) for various types of two Higgs doublet models in the alignment limit to find the range of $u_{H}$. Next we fix $u_{H}= 0.01c$ and find the range of $\tan\beta$. The results are displayed in Table~\ref{Relative velocity tan beta.Table.1}. In \textit{Natural System of Units} we set $c=1$ and thus we will refrain from writing $c$ from now onwards.
\begin{table}[htb]
\begin{tabular}{||c|c|c|c||}
\hline
2HDMs & $\xi_{H}^{b}\; $
&\; Limit of $u_{H}$ &\; Limit of $\tan\beta$\\
&&\; for $\tan\beta=$ 5&\; for $u_{H}=$0.01\\
\hline
Type I &$\;-\cot\beta $ 
&$\;  u_{H} > \mathrm{0.08}$ &$\; \tan\beta>$39.89\\
\hline 

Type II &$\;\tan\beta $ 
&$\;  u_{H} > \mathrm{1.99}$ &$\;\tan\beta<$ 0.025\\
\hline 
Lepton Specific &$\;-\cot\beta $ 
&$\;  u_{H} > \mathrm{0.08}$ &$\; \tan\beta>$39.89\\
\hline 
Flipped &$\;\tan\beta $ 
&$\;  u_{H} > \mathrm{1.99}$ &$\;\tan

\beta<$ 0.025\\
\hline
\end{tabular}
\caption{Limits for relative velocity for $\tan\beta = 5$ and $\tan\beta$ for $u_{H}=0.01$ when $m_{h}<m_{H}< 2m_{t}$.}
\label{Relative velocity tan beta.Table.1}
\end{table}
%
We see that when we fix $\tan\beta=5\,,$  the value calculated for $u_{H}$ for the type II and flipped 2HDMs
is not sensible. Similarly, when we set $u_H=0.01$\,, the bound on $\tan\beta$ is far too low. Thus we conclude that
these two types of 2HDMs do not seem to allow the formation of $H-H$ bound states. In the other two types of 
2HDMs also, the formation of $H-H$ bound state is not very easy, since $\tan\beta$ and $u_{H}$ both must take 
rather high values, and thus are at the edge of the region of validity for the non-relativistic analysis used here.

\subsubsection*{Case II: $m_{H}> 2m_{t}$}
In the case of a heavy $H$ particle with  $m_{H}> 2m_{t}$\,,  the predominant decay channels are 
$H \rightarrow b \overline{b}$ and $H \rightarrow t \overline{t}$\,. Neglecting the effects of new physics operators we can 
calculate the decay width of $H$ into $t \overline{t}$ pair as
\begin{equation}
\Gamma_{t}^{H} = \frac{m_{H}(\xi_{H}^{t})^{2}}{8\pi}(1-4\frac{m_{t}^{2}}{m_{H}^{2}})^{3/2}\,. \label{gamma_tH}
\end{equation}
The total decay width is then approximately $\Gamma^{H} = \Gamma^{H}_{b}+\Gamma^{H}_{t}\,$ where the expression for 
$\Gamma^{H}_{b}$ is given in Eq.~(\ref{gamma_bH}), and the decay time is the inverse of the total decay width, 
$\tau^{H}=(\Gamma^{H})^{-1}$\,.

The range of the mass of $H$ in the \textit{Alignment limit} was derived in~\cite{Biswas:2014uba} 
as $m_{H}\in [450,620]$ GeV.
Here we estimate $\tau^{H}$ for the two extreme values of $m_H$ in this range, namely $m_{H}= $450 
and 620 GeV for various types of 2HDMs in the alignment limit and for $\tan \beta=5$\,. We also display the lower limits of the relative velocity using the logic that the formation time of the bound state must be shorter than the decay time of the 
parent particle if the bound state is to form. For all types of 
2HDMs in alignment limit the relative $H t\bar{t}$ coupling is $\xi_{H}^{t} = -\cot\beta$  and as we have seen in the first case $\xi_{H}^{b}$ is type dependent. The results are displayed in the Table~\ref{Decay time.Table.2}. 
\begin{table}[htbp]
\begin{tabular}{||c|c|c|c|c||}
\hline
2HDMs &$\;  \tau^{H}$ in secs  &$\; u_{H}>\frac{2}{\sqrt{m_{H}\tau^{H}}}$ &$\;  \tau^{H}$ in secs  &$\; u_{H}>\frac{2}{\sqrt{m_{H}\tau^{H}}}$\\
&{\small ($m_{H}$=450 GeV)}&&{\small ($m_{H}$=620 GeV)}&\\
\hline
Type I  & 7.25$\times \mathrm{10^{-25}}$ &$u_{H}>$0.09   &  4.24$\times \mathrm{10^{-25}}$ &$\;u_{H}>$0.1\\
\hline 
Type II  & 1.47$\times \mathrm{10^{-27}}$ &$\;u_{H}>$1.99 &  1.06$\times \mathrm{10^{-27}}$ &$\;u_{H}>$1.99\\
\hline 
Lepton Specific & 7.25$\times \mathrm{10^{-25}}$ &$u_{H}>$0.09   &  4.24$\times \mathrm{10^{-25}}$ &$\;u_{H}>$0.1\\
\hline 
Flipped & 1.47$\times \mathrm{10^{-27}}$ &$\;u_{H}>$1.99 &  1.06$\times \mathrm{10^{-27}}$ &$\;u_{H}>$1.99\\
\hline
\end{tabular}
\caption{Decay time (in seconds) and relative velocity for $m_{H}= 450$ GeV and 620 GeV  and $\tan\beta = 5$ for the different 2HDMs.}
\label{Decay time.Table.2}
\end{table}

For Type I and 
lepton specific models, $u_{H}$ can still be said to be in the non-relativistic range, but for Type II and flipped 
models, the value of $u_{H}$ is not sensible. If we choose a non-relativistic value of $u_{H}$\,, say $u_{H}=0.01$\,, 
it follows that $\tan\beta$ takes values greater than 56.4 for Type I and lepton specific models, whereas for 
Type II and flipped models $\tan\beta$ takes no permissible value. 

It is clear that $H-H$ bound state will not form in the Type II and flipped 2HDMs, but may form in 
Type I and lepton specific models. But even that conclusion is not a strong one, as the range of
parameters for bound state formation are at the edge of the allowed values. 
\section{Summary and Conclusion}
In this work we have applied the non-relativistic version of Higgs effective field theory to estimate the existence of bound states of the heavier CP-even Higgs boson of 2HDMs. The relative strengths of the attractive and repulsive contact interactions were determined and the possibility of the formation of bound state was predicted. It was found that for obese Higgs the attractive coupling was stronger than the repulsive coupling thus facilitating the formation of the bound state. Moreover as $m_{H}$ increased the possibility was high. Next we have considered the decay of the parent particle into various channels and the corresponding decay times were estimated. Decay times were compared with the formation time of the bound state. If the formation time was smaller than the decay time, then bound state formation would take place readily. From this approach we concluded that for obese Higgs the bound states are likely to form in Type - I and lepton specific models but not in Type -II and flipped 2HDMs.\\
 
Various works on the bound states of Higgs bosons have shown that only obese Higgs bosons have the tendency to form bound states. The peak at 125 GeV has forced us to identify it with the predicted Higgs boson and eventually for two Higgs doublet models the most reasonable thing to do would be to stay in the alignment limit assigning $m_{h} = 125$ GeV. For this very reason we have studied the bound state formation of the CP even non-standard Higgs boson whose mass spectra is flexible.  Though we have imposed Naturalness conditions and have restricted the mass of $H$ within bounds, still these bounds are much higher on the mass scale and make the probability of bound state formation a possibility. If the Naturalness criteria is withdrawn and the potential is only subjected to stability and perturbative unitarity constraints then $m_{H}$  is much more flexible and the entire spectrum can be studied for the possibility of the bound state formation.
\\
 
There are many questions that we have not addressed. The immediate one is the mass of the bound state and its life time. Another important point is the detectability of the bound state. In future works we could attempt to address these questions.\\

In future we would like to study the possibility of formation of h-H bound state. Further since the $\lambda_{i}'s$ get restricted when expressed in terms of the masses of the physical Higgs bosons of the two Higgs doublet model, we would like to study the variation of effective cubic (attractive) and quartic (repulsive) coupling strengths with $\lambda_{i}'s$ maintaining the perturbative unitarity condition. We also intend to study the formation of bound states by elevating the restrictions imposed by \textit{Naturalness}. It would also be interesting to solve the bound state equation in the more general, fully relativistic case.

\appendix
\section{CUSTODIAL SYMMETRY}\label{Appendix A}

In ~\cite{Willenbroc:2004} the two Higgs doublet fields are given as

\begin{equation}
\phi_{i}=
\left(
\begin{array}{c}
\phi_{i}^{+}\\
\phi_{i}^{0}
\end{array}
\right)\,, \qquad i=1,2.
\end{equation}

Then $\epsilon\phi_{i}^{\star}$ are also two $SU(2)_{L}$ doublets with components

\begin{equation}
\epsilon\phi_{i}^{\star}=
\left(
\begin{array}{c}
\phi_{i}^{0\star}\\
-\phi_{i}^{-}
\end{array}
\right)\,, 
\end{equation}

where $\phi_{i}^{-} = \phi_{i}^{+\star}$. The Higgs bi-doublet fields are given by

\begin{eqnarray}
\Phi_{i} &=&
\frac{1}{\sqrt{2}}
\left(
\begin{array}{rc}
\epsilon\phi_{i}^{\star}& \phi_{i}
\end{array}
\right)\,, \nonumber\\
&=& 
\frac{1}{\sqrt{2}}
\left(
\begin{array}{rc}
\phi_{i}^{0 \star}& \phi_{i}^{+}\\
-\phi_{i}^{-}& \phi_{i}^{0}
\end{array}
\right)\,.
\end{eqnarray}

The $SU(2)_{L} \times U(1)_{Y}$ gauge symmetry acts on the Higgs bi-doublets as
\begin{eqnarray}
SU(2)_{L} &:& \Phi_{i}\rightarrow L \Phi_{i}\\
U(1)_{Y} &:& \Phi_{i}\rightarrow  \Phi_{i}e^{-i\sigma_{3}\theta_{i}/2}\,.
\end{eqnarray}
The Lagrangian has the following global symmetry in the limit where hypercharge vanishes
\begin{equation}
SU(2)_{R} : \Phi_{i}\rightarrow  \Phi_{i} R^{\dagger}\,.
\end{equation}
When the Higgs fields acquire their respective vacuum expectation values, both $SU(2)_{L}$ and $SU(2)_{R}$ are broken, however the subgroup $SU(2)_{L=R}$ is unbroken, i.e at $\left\langle\phi_{i}^{0}\right\rangle =v_{i}\,,$ one has
\begin{equation}
\left\langle \Phi_{i}\right\rangle = \frac{1}{\sqrt{2}}
\left(
\begin{array}{rc}
v_{i}^{\star}& 0\\
0& v_{i}
\end{array}
\right)\,.
\end{equation}
When the vacuum expectation values are chosen to be real, $v_{i}^{\star}=v_{i}$, $\left\langle \Phi_{i}\right\rangle $ is proportional to the $2\times 2$ identity matrix and the vacuum preserves a group $SU(2)_{V}$ (the $V$ stands for ``vectorial") corresponding to the identical matrices $SU(2)_{L=R}$ i.e,
\begin{equation}
L\left\langle\Phi_{i}\right\rangle L^{\dagger}=\left\langle\Phi_{i}\right\rangle\,.
\end{equation}
This remaining group preserved by the vacuum is the custodial-symmetry group.





\section*{Acknowledgement}

I would like to thank my supervisor Prof. Amitabha Lahiri for his constant help in writing this paper. I would also like to thank Dr. Debasis Mukherjee, Dr. Jayeeta Saha and Tushar Kanti Biswas for useful discussion on the topic.

\end{document}